\documentclass{siamltex}
\usepackage{bm}
\usepackage{bbm}
\usepackage{graphicx}
\usepackage{url}
\usepackage{color}
\usepackage{mathtools}
\usepackage{amsmath}	
\usepackage{amssymb}	
\usepackage{graphicx,graphics,color}	
\usepackage{enumerate}
\usepackage{cancel}
\usepackage{color}
\usepackage{epsfig}

\newtheorem{thm}{Theorem}[section] 
\newtheorem{remark}[thm]{Remark}

\usepackage{algorithmic}	
\usepackage[section]{algorithm}

\title{Pattern formation and oscillations in nonlinear random walks on networks
\thanks{
This work was funded in part by NSF Grant No.~MCB-2126177 (PSS). Any opinions, findings, and conclusions or recommendations expressed in this material are those of the author(s) and do not necessarily reflect the views of the funding agencies. 
}}
\author{
Per Sebastian Skardal\thanks{Department of Mathematics, Trinity College, Hartford, CT, 06106, USA}
}

\graphicspath{{figures/}}

\begin{document}

\maketitle

\begin{abstract}
Random walks represent an important tool for probing the structural and dynamical properties of networks and modeling transport and diffusion processes on networks. However, when individuals' movement becomes dictated by more complicated factors, e.g., scenarios that involve complex decision making, the linear paradigm of classical random walks lack the ability to capture dynamically rich behaviors. One modification that addresses this issue is to allow transition probabilities to depend on the current system state, resulting in a {\it nonlinear random walk}. While the resulting nonlinearity has been shown to give rise to an array of more complex dynamics, the patterns that emerge, in particular on regular network topologies, remain unexplored and poorly understood. Here we study nonlinear random walks on regular networks. We present a number of stability results for the uniform state where random walkers are uniformly distributed throughout the network, characterizing the spectral properties of its Jacobian which we use to characterize its bifurcations. These spectral properties may also be used to understand the patterns that emerge beyond bifurcations, which consist of oscillating short wave-length patterns and localized structures for negative and positive bias, respectively. We also uncover a subcriticality in the bifurcation for positive bias, leading to a hysteresis loop and multistability.
\end{abstract}

\begin{keywords}
{Nonlinear random walks, pattern formation, complex networks}
\end{keywords}

\begin{AMS}
05C81, 60J10, 60J20, 60J60, 39A06, 37N20
\end{AMS}

\pagestyle{myheadings}
\thispagestyle{plain}
\markboth{P. S. Skardal}{Pattern formation in nonlinear random walks}

\section{Introduction}\label{sec:01}

The dynamics of random walks and diffusion processes have long been utilized for their ability to shed light on the structural properties of and the dynamics on complex networks~\cite{Masuda2017PhysRep}. For instance, random walks serve as the foundation of Google's PageRank algorithm for ranking node centralities~\cite{Gleich2015SIAM} and may be used for identifying mesoscopic structure such as communities~\cite{Rosvall2008PNAS} and geometric embeddings~\cite{Coifman2005PNAS} of networks, while their dynamics have been used for network exploration~\cite{Sinatra2011PRE,Battiston2016NJP} and modeling transport processes~\cite{Nicosia2017PRL,Cencetti2018PRE}. The dynamics of classical discrete-time random walks on finite networks are relatively simple, as the evolution of occupation probabilities of random walkers throughout the network are described by a linear dynamical system. These simple dynamics, even in the face of large, topologically complex network structures are key to their widespread utility and flexibility across applications.

From the point of modeling transport processes, however, the simplicity of classical random walks may be overly constraining. Take for example the transport of resources between various institutions. Depending on the properties and goals of the institutions, decisions that dictate where resources are moved to and invested may be much more complex. For instance, humanitarian values may drive resources towards relatively poor institutions, whereas more capitalistic values may incentivize investment into already wealthier institutions. This and other transport processes that depend on complex decision-making, such as travel and immigration, are likely complicated enough so that the linear dynamics of a classic random walk is insufficient to realistically capture important behaviors.

One generalization that addresses this important lack of complexity is to allow the transition probabilities to be non-static, and rather allow them to depend on the current state of the system, i.e., the occupation probabilities of the nodes in the network. This results in a random walk that evolves according to a nonlinear dynamical system and therefore may display much more complicated dynamical behaviors such as periodic solutions, quasi-periodic solutions, multistability, and chaos~\cite{Skardal2019JNS,Skardal2020PRE,Carletti2020PRR,Skardal2021PRR}. While these dynamics have been investigated primarily in the context of complex, heterogeneous networks structures, open questions remain in understanding, as with any diffusion process, their behavior on more regular topologies and the patterns that form on these topologies.

In this paper we investigate the dynamics of nonlinear random walks on networks with homogeneous structure, and in particular the pattern formation and onset of oscillations that emerge. In particular, we show that when placed on a regular network topology, the uniform state, where random walkers are distributed uniformly throughout the network, is stable provided that the nonlinearity driving random walk bias is not too strong. However, this uniform state undergoes a pair of local bifurcations if the transition bias is sufficiently positive or negative, giving rise to localized or oscillating patterns, respectively. We show that the eigenvalue spectrum of the Jacobian matrix of the uniform state, which dictates the stability of the uniform state itself, is a nonlinear function of the eigenvalue spectrum of the network adjacency matrix, then use these theoretical results to explore the bifurcations of the uniform state and the pattern formation phenomena that occurs beyond these bifurcations. In particular, we consider both ring and small-world~\cite{Watts1998Nature}.

The remainder of this paper is organized as follows. In section~\ref{sec:02} we present some mathematical preliminaries of classical random walks on networks, introduce nonlinear random walks on networks, and present some motivating examples of pattern formation in nonlinear random walks. In section~\ref{sec:03} we study the stability and bifurcations of the uniform state. In particular, we present a series of theoretical results that first demonstrate that the uniform state is always a fixed point for a regular network, then links the eigenvalue spectrum of the Jacobian of the uniform state to the eigenvalue spectrum of the adjacency matrix. We then explore the stability of the uniform state and its bifurcations in classes of ring and small-world networks. In section~\ref{sec:04} we explore the patterns that form beyond these bifurcations, which may be understood using the dominant eigenvectors of the Jacobian matrix of the uniform state. In section~\ref{sec:05} we explore hysteresis in the case of positive bias. Finally, in section~\ref{sec:06} we conclude with a discussion of our results.

Random walks review~\cite{Maccluer2000SIAM}

\section{Preliminaries and Nonlinear Random Walks}\label{sec:02}

We first present some mathematical preliminaries of linear random walks on networks (subsection~\ref{sec:02:01}). Next we introduce nonlinear random walks (subsection~\ref{sec:02:02}). Lastly, we present some motivating examples of pattern formation in nonlinear random walks on networks (subsection~\ref{sec:02:03}).

\subsection{Preliminaries: Linear Random Walks}\label{sec:02:01}

We begin by discussing the basic properties classical discrete-time random walks on finite networks, of which nonlinear random walks are a generalization.

\begin{definition}[Discrete-Time Random Walks on Finite Networks]\label{def:02:01}
Consider a network of $N\in\mathbb{N}$ nodes encoded by the adjacency matrix $A$, whose entries $A_{ij}$ describe the presence and strength of links from a node $j$ to another node $i$, with $i,j = 1,\dots,N$.
Denote the probability of a random walker occupying node $i$ at time $t$ as $p_i(t)$. Then the occupation probabilities evolve according to
\begin{align}
p_i(t+1) = \sum_{j = 1}^N \pi_{ij}p_j(t),\label{eq:02:01}
\end{align}
where $\pi_{ij}=A_{ij}/k^{\text{out}}_{j}$ is the (conditional) transition probability of moving from node $j$ to node $i$ in one time step and $k^{\text{out}}_{j}=\sum_{i=1}^NA_{ij}$ is the out-degree of node $j$. Note that by storing the occupation probabilities in the vectors $\bm{p}(t)\in[0,1]^N$ and the transition probabilities in the matrix $\Pi\in[0,1]^{N\times N}$, \eqref{eq:02:01} may be written in vector form as
\begin{align}
\bm{p}(t+1) = \Pi\bm{p}(t).\label{eq:02:02}
\end{align}
\end{definition}

\begin{remark}\label{rem:02:01}
In general we assume that the entries of the adjacency matrix are non-negative. Here we will focus on the undirected, unweighted case were $A_{ij}=A_{ji}$ (i.e., $A=A^T$) and $A_{ij}=1$ if a link exists between nodes $i$ and $j$ and otherwise $A_{ij}=0$, but these constraints may be relaxed to allow a network to be directed or weighted.
\end{remark}

\begin{remark}\label{rem:02:02}
In Def.~\ref{def:02:01} the occupation probability vector $\bm{p}(t)$ must sum to one, i.e., $\sum_{i=1}^Np_i(t)=1$. Moreover, the transition matrix $\Pi$ contains conditional probabilities, and therefore is column-stochastic, i.e., $\sum_{i=1}^N\pi_{ij} = 1$.
\end{remark}

\begin{remark}\label{rem:02:03}
In Def.~\ref{def:02:01} the transition probabilities are normalized by the out-degree of each node, yielding an {\it unbiased}, a.k.a., classical, random walk. Alternatively, a {\it biased} random walk breaks the uniformity of the non-zero entries of each column of the transition matrix $\Pi$. This may be achieved incorporating a vector $f_i$ into the definition of the transition probabilities, namely, $\pi_{ij}=A_{ij}f_i/\left(\sum_{l=1}^NA_{lj}f_l\right)$, to direct random walkers towards or away from nodes according to the values encoded in the vector $\bm{f}$~\cite{Gomez2008PRE,Bonaventura2014PRE}.
\end{remark}

The classical random walks (both unbiased and biased) given in Def.~\ref{def:02:01} have several key properties that make them particularly useful for a wide range of applications. First and foremost, whether the random walk is unbiased or biased, the static nature of the transition probabilities ensure that \eqref{eq:02:02} constitutes a {\it linear} dynamical system that is stepped forward in time by a simple matrix multiplication. Second, this linearity ensures that, under the relatively mild conditions of the underlying network structure being primitive, i.e., there exists some positive integer $m$ such that the matrix $\Pi^m$ has all (strictly) positive entries, there exists a unique, globally-attracting fixed point $\bm{p}^*$, a.k.a. stationary distribution. (Note that the network being primitive is equivalent to the Markov chain being irreducible and aperiodic.)

These properties yield simple dynamics, regardless of the size or structure of the underlying network, assuring that random walks on networks find a wide range of applications. This is due to the linearity of the system in \eqref{eq:02:02}, which itself follows from the static nature of the transition probabilities. Despite the mathematical advantages of static transition probabilities in a random walk, this assumption may be unrealistic from a modeling perspective, for instance when used to describe a transport process.

\subsection{Nonlinear Random Walks}\label{sec:02:02}

We now define a nonlinear random walk on a network as a generalization of classical random walks where transition probabilities are not static, and rather depend on the current state of the system, i.e., the occupation probabilities at different nodes.

\begin{definition}[Nonlinear Random Walks on Networks]\label{def:02:02}
Consider a network of $N$ nodes with adjacency matrix $A$. Next, denote the bias function $f:[0,1]\to(0,\infty)$. Then the nonlinear random walk induced by the bias function $f$ on the network with adjacency matrix $A$ evolves according to
\begin{align}
p_i(t+1) = \sum_{j = 1}^N \pi_{ij}(\bm{p}(t))p_j(t),\label{eq:02:03}
\end{align}
where the transition probabilities are now given by
\begin{align}
\pi_{ij}(t) =\frac{A_{ij}f(p_i(t))}{\sum_{l=1}^NA_{lj}f(p_l(t))}.\label{eq:02:04}
\end{align}
In vector form, \eqref{eq:02:03} may be written as
\begin{align}
\bm{p}(t+1) = \Pi(\bm{p}(t))\bm{p}(t).\label{eq:02:05}
\end{align}
\end{definition}

\begin{remark}\label{rem:02:05}
The nonlinear random walk defined above is best used in models involving a continuum of random walkers. Moreover, the functional form of the bias function $f$ plays a critical role in shaping the nonlinear random walk in Def.~\ref{def:02:02}. Namely, if $f$ is monotonically increasing (decreasing) then walkers are preferentially directed towards nodes that are currently occupied by a relatively large (small) fraction of walkers. In the remainder of this work we consider the choice $f(p)=\text{exp}(\alpha p/z)$, where $z=N^{-1}$, yielding transition probabilities
\begin{align}
\pi_{ij}(\bm{p}(t)) =\frac{A_{ij}\text{exp}(\alpha p_i(t)/z)}{\sum_{l=1}^NA_{lj}\text{exp}(\alpha p_l(t)/z)},\label{eq:02:06}
\end{align}
where positive (negative) values of the parameter $\alpha$ bias walks towards nodes with currently high (low) occupancy. The value $z=N^{-1}$ is chosen to ensure that dynamics remain similar across different network sizes.
\end{remark}

The introduction of state-dependent transition probabilities yield a dynamical system in \eqref{eq:02:03} that is nonlinear. In weakly nonlinear regimes (e.g., small $|\alpha|$ for the exponential bias function choice described above) it has been shown that the dynamics relax to, as in the case of the standard random walk with static transition probabilities, a unique, globally-attracting stationary distribution, assuming the same primitiveness condition on the network structure is met~\cite{Skardal2019JNS}. However, under stronger nonlinearity more complicated dynamics occur, including multistability, periodic solutions, quasi-periodic, and even chaotic states when the bias function is non-monotonic~\cite{Skardal2020PRE,Skardal2021PRR}. Moreover, in the continuous-time analogue nonlinear random walks have been shown to optimize network exploration~\cite{Carletti2020PRR}. These dynamical behavior have been investigated primarily in the context of heterogeneous network structures, leaving an understanding of the behavior of nonlinear random walks on more homogeneous topologies lacking. Therefore, in this paper we focus our attention on the dynamics of pattern formation on undirected, regular network structures, i.e., networks where all nodes have the same nodal degree. 

\subsection{Pattern Formation and Oscillations in Nonlinear Random Walks}\label{sec:02:03}

\begin{figure}[t]
\centering
\epsfig{file =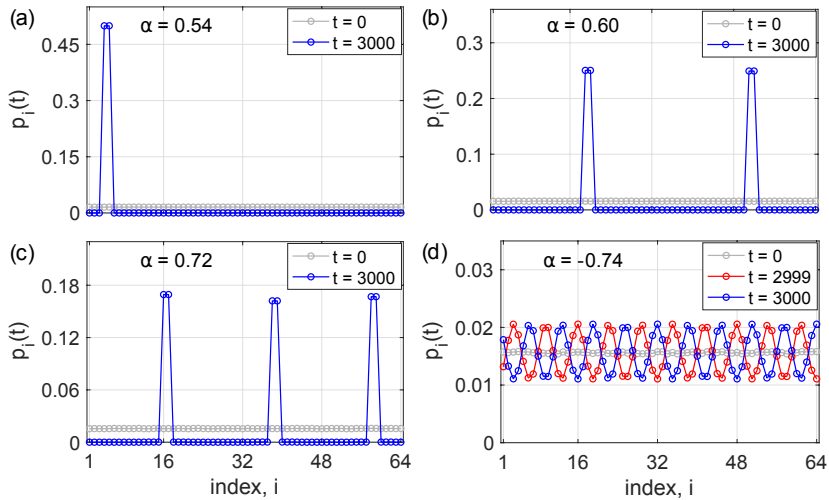, clip =,width=0.85\linewidth }
\caption{{\it Pattern formation in nonlinear random walks.} Example patterns for a nonlinear random walk on a ring network of $N=64$ nodes with connectivity radius $r=4$. Localized patterns with one, two, and three localized spots form for positive bias (a) $\alpha=0.54$, (b) $0.60$, and (c) $0.72$, respectively. Oscillating waves form for negative bias (d) $\alpha = -0.74$.} \label{fig1}
\end{figure}

Next we present some motivating examples of pattern formation in nonlinear random walks on networks. In particular, we consider a family of ring networks characterized by a connectivity radius $r$, such that each node is connected to the $r$ nodes on either side, thereby resulting in a regular network where each node has degree equal to twice the connectivity radius, i.e., $k=2r$. Taking a network of $N=64$ nodes and connectivity radius $r=4$ we then simulate the nonlinear random walk dynamics described by \eqref{eq:02:03} and \eqref{eq:02:06}. For initial conditions we consider a state that is randomly perturbed from the uniform state, $p_i(0) = 1/N + \delta p_i$, where the perturbation vector $\bm{\delta p}$ has zero mean and norm $\|\bm{\delta p}\|=10^{-3}$. In Fig.~\ref{fig1} we present the results for four different choices of $\alpha$, namely, $\alpha=0.54$, $0.60$, $0.72$, and $-0.74$ in (a)--(d), plotting the initial state $\bm{p}(0)$ in light grey circles and the steady-state obtained after $3000$ time steps in blue circles. Beginning with $\alpha=0.54$ (a) we find a steady-state solution consisting of a single localized structure spanning two adjacent nodes where essentially all the random walkers end up. When $\alpha$ is increased to $0.60$ (b) and $0.72$ (c) we see similar results, except for he fact that the steady state is made up of not one, but two and three localized structures, respectively. Finally, for a sufficiently negative bias of $\alpha = -0.74$ (d), the resulting pattern is a wave-like patter with relatively short spatial wavelength that is also oscillating in time, as we have plotted the state at the prior time step in red circles.

These examples highlight a number of features of the patterns that form in nonlinear random walks on rings and homogeneous networks in general. First, for positive $\alpha$, i.e., when random walkers are biased towards nodes where there are already relatively many random walkers, the resulting patterns are highly localized. As we will see, the properties of this localization, in particular the number of localized structures that emerge, depends on $\alpha$ can be determined by the spectral properties of the uniform state. These patterns form after a transcritical bifurcation that turns out to be subcritical. On the other hand, for negative $\alpha$, i.e., when random walkers are biased towards nodes where there are currently relatively few random walkers, the resulting patterns that form are wave-like with a clearly observable spatial wavelength. Moreover, the loss of stability that results in these wave patterns occurs as a period-doubling bifurcation, as the wave pattern itself undergoes a period-two temporal oscillation. These dynamics can also be explained by studying the spectral properties of the uniform state.

\section{Stability and Bifurcations of the Uniform State}\label{sec:03}

Our analysis of pattern formation and oscillations in nonlinear random walks begins with the uniform state. For the remainder of this paper we will assume, unless otherwise stated, that networks are both undirected and regular, i.e., the adjacency matrix $A$ of the network satisfies $A_{ij}=A_{ji}$ for all $i,j=1,\dots,N$ and all nodes have the same degree, that is $k=k_i^{\text{in}}=k_i^{\text{out}}$ for all $i$, where $k_i^{\text{in}}=\sum_{j=1}^NA_{ij}$ and $k_j^{\text{out}}=\sum_{i=1}^NA_{ij}$ denote in- and out-degrees. We begin by presenting a number of stability results for the uniform state (subsection~\ref{sec:03:01}), which we use to explore the bifurcations in the system dynamics (subsection~\ref{sec:03:02}).

\subsection{Stability of the Uniform State}\label{sec:03:01}

We begin by analyzing the uniform state where all occupation probabilities throughout the network are identical, i.e., $p_i(t)=1/N$ for all $i=1,\dots,N$, or in vector notation, $\bm{p}=\bm{1}/N$. First we show that, regardless of the choice of bias function $f$, the uniform state is a fixed point of the system provided that the network is regular.

\begin{proposition}[Uniform States are Fixed Points]\label{prop:03:01}
Consider a nonlinear random walk defined in Def.~\ref{def:02:02} on a regular network. Then the uniform state, $\bm{p}_{\text{unif}}=\bm{1}/N$, is a fixed point for any appropriately chosen bias function $f$.
\begin{proof}
Let $p_i(t)=1/N$ for all $i=1,\dots,N$. Then $p_i(t+1)$ is given by
\begin{align}
p_i(t+1) &= \sum_{j = 1}^N \frac{A_{ij}f(p_i(t))}{\sum_{l=1}^NA_{lj}f(p_l(t))}p_j(t)\\
&= \sum_{j = 1}^N \frac{A_{ij}f(1/N)}{\sum_{l=1}^NA_{lj}f(1/N)}\frac{1}{N}\\
&=  \frac{1}{N} \sum_{j = 1}^N \frac{A_{ij}}{k}= \frac{1}{N},
\end{align}
where we've used that $\sum_{l=1}^NA_{lj}=k$. This concludes the proof.
\end{proof}
\end{proposition}

\begin{remark}\label{rem:03:01}
Note that Proposition~\ref{prop:03:01} does not require the network structure to be undirected, as it hold for undirected network, as long as the network is regular.
\end{remark}

Next we turn our attention to the Jacobian matrix of the system, which dictates that stability of the uniform state. We first show that, when evaluated at the uniform state, the Jacobian is a simple but nonlinear function of the adjacency matrix $A$, the degree $k$, and the bias parameter $\alpha$.

\begin{lemma}[Jacobian of the Uniform State]\label{lem:03:02}
Consider a nonlinear random walk defined in Def.~\ref{def:02:02} with exponential bias function $f$ on an undirected, regular network. Then the Jacobian matrix $DF$ evaluated at the uniform state $\bm{p}_{\text{unif}}=\bm{1}/N$ is given by
\begin{align}
DF(\bm{p}_{\text{unif}}) = \frac{A}{k}-\alpha \frac{A^2}{k^2}+\alpha I
\end{align}
\begin{proof}
First, a straight-forward calculation yields that the entries of the Jacobian are given by
\begin{align}
DF_{ij}(\bm{p}) &= \frac{A_{ij}\text{exp}(\alpha p_i/z)}{\sum_{l=1}^NA_{lj}\text{exp}(\alpha p_l/z)} - \frac{\alpha}{z}\sum_{m=1}^N\left(\frac{A_{im}A_{jm}\text{exp}(\alpha p_i/z)\text{exp}(\alpha p_j/z)}{\left(\sum_{l=1}^NA_{lj}\text{exp}(\alpha p_l/z)\right)^2}\right)p_m\\
&~~~~+\delta_{ij}\frac{\alpha}{z}\sum_{m=1}^N\left(\frac{A_{im}\text{exp}(\alpha p_i/z)}{\sum_{l=1}^NA_{lm}\text{exp}(\alpha p_l/z)}\right)p_m,
\end{align}
where $\delta_{ij}$ is the Kronecker delta function. Next, note that for the uniform state, $p_i/z=(1/N)/(1/N)=1$. Together this allows us to write
\begin{align}
DF_{ij}(\bm{p}_{\text{unif}}) &= \frac{A_{ij}\text{exp}(\alpha)}{\sum_{l=1}^NA_{lj}\text{exp}(\alpha)} - \frac{\alpha}{z}\sum_{m=1}^N\left(\frac{A_{im}A_{jm}\text{exp}(\alpha)\text{exp}(\alpha)}{\left(\sum_{l=1}^NA_{lj}\text{exp}(\alpha)\right)^2}\right)\frac{1}{N}\\
&~~~~+\delta_{ij}\frac{\alpha}{z}\sum_{m=1}^N\left(\frac{A_{im}\text{exp}(\alpha)}{\sum_{l=1}^NA_{lm}\text{exp}(\alpha)}\right)\frac{1}{N}\\
&= \frac{A_{ij}}{\sum_{l=1}^NA_{lj}} - \alpha\sum_{m=1}^N\frac{A_{im}A_{jm}}{\left(\sum_{l=1}^NA_{lj}\right)^2}+\delta_{ij}\alpha\sum_{m=1}^N\frac{A_{im}}{\sum_{l=1}^NA_{lm}}\\
&= \frac{A_{ij}}{k} - \frac{\alpha}{k^2}\sum_{m=1}^NA_{im}A_{jm}+\delta_{ij}\frac{\alpha}{k}\sum_{m=1}^NA_{im}\\
&=\frac{A_{ij}}{k} - \frac{\alpha}{k^2}\sum_{m=1}^NA_{im}A_{mj}+\delta_{ij}\alpha,
\end{align}
where we have used the undirectedness of the network to replace $A_{jm}$ with $A_{mj}$. Noting that $\sum_{m=1}^NA_{im}A_{mj} = (A^2)_{ij}$, we obtain $DF(\bm{p}_{unif})=A/k-\alpha A^2 I + \alpha I$, which completes the proof.
\end{proof}
\end{lemma}

Next we use Lemma~\ref{lem:03:02} to link the spectrum of the Jacobian of the uniform state to that of the adjacency matrix $A$. In particular, the eigenvector spectrum of the Jacobian is precisely equal to the eigenvector spectrum of the adjacency matrix, and its eigenvalues are a nonlinear function of the eigenvalues of the adjacency matrix, the degree, and the bias parameter.

\begin{theorem}[Spectrum of the Uniform State]\label{thm:03:03}
Consider a nonlinear random walk defined in Def.~\ref{def:02:02} with exponential bias function $f$ on an undirected, regular network. Then if $\bm{v}\in\mathbb{R}^N$ is an eigenvector of the adjacency matrix $A$ with corresponding eigenvalue $\lambda_A$, $\bm{v}$ is also an eigenvector of the Jacobian matrix evaluated at the uniform state, denoted $DF_{\text{unif}}=DF(\bm{p}_{\text{unif}})$, and its corresponding eigenvalue is
\begin{align}
\lambda_{DF}=\frac{\lambda_A}{k} - \alpha\frac{\lambda_A^2}{k^2} +\alpha.
\end{align}
\begin{proof}
From Lemma~\ref{lem:03:02} we have first that $DF_{unif}=A/k-\alpha A^2 I + \alpha I$. Now let $\bm{v}$ be an eigenvector of $A$ with eigenvalue $\lambda_A$ so that $A\bm{v}=\lambda_A\bm{v}$. Multiplication of $\bm{v}$ by $DF_{\text{unif}}$ then yields
\begin{align}
DF_{\text{unif}}\bm{v}&=\left(\frac{A}{k}-\alpha \frac{A^2}{k^2}+\alpha I\right)\bm{v}\\
&=\frac{A\bm{v}}{k}-\alpha \frac{A^2\bm{v}}{k^2}+\alpha\bm{v}\\
&=\frac{\lambda_A\bm{v}}{k}-\alpha \frac{\lambda_A^2\bm{v}}{k^2}+\alpha\bm{v}\\
&=\left(\frac{\lambda_A}{k}-\alpha \frac{\lambda_A^2}{k^2}+\alpha\right)\bm{v},
\end{align}
which completes the proof.
\end{proof}
\end{theorem}

\begin{remark}\label{rem:03:02}
In Theorem~\ref{thm:03:03} we have denoted the eigenvalues of the Jacobian $DF_{\text{unif}}$ and the adjacency matrix $A$ as $\lambda_{DF}$ and $\lambda_A$, respectively, in order to differentiate the two.
\end{remark}

Finally, our interest in the eigenvalue spectrum of the uniform state stems from its ability to inform the stability of the uniform state to perturbation, for which we now define the dominant eigenvalue. However, nonlinear random walks lie on the center manifold where probability is conserved, i.e., the sum $\sum_{i=1}^Np_i(t)$ is a conserved quantity. This center manifold is characterized mathematically by a trivial eigenvalue $\lambda=1$ of the Jacobian.

\begin{proposition}[Trivial eigenvalue of the Jacobian]\label{prop:03:04}
Consider a nonlinear random walk defined in Def.~\ref{def:02:02} with exponential bias function $f$ on an undirected, regular network. Then the Jacobian matrix evaluated at the uniform state $DF_{\text{unif}}$ has a trivial eigenvalue $\lambda=1$ with a corresponding eigenvector that is constant, $\bm{v}\propto\bm{1}$.
\begin{proof}
From Theorem~\ref{thm:03:03} we have first that $DF_{unif}=A/k-\alpha A^2 I + \alpha I$. Upon matrix multiplication of the constant vector $\bm{1}$ we have that
\begin{align}
DF_{\text{unif}}\bm{1}&=\left(\frac{A}{k}-\alpha \frac{A^2}{k^2}+\alpha I\right)\bm{1}\\
&=\frac{A\bm{1}}{k}-\alpha \frac{A^2\bm{1}}{k^2}+\alpha\bm{1}\\
&=\frac{\bm{k}}{k}-\alpha \frac{\bm{k}^2}{k^2}+\alpha\bm{k}\\
&=\bm{1} - \alpha\bm{1} + \alpha\bm{1}=\bm{1},
\end{align}
where we have used that $A\bm{1}=\bm{k}$ and $A^2\bm{1}=\bm{k^2}$, where $\bm{k}=[k,k,\dots,k]^T$ and $\bm{k^2}=[k^2,k^2,\dots,k^2]^T$. This completes the proof.
\end{proof}
\end{proposition}

Given the existence of this trivial eigenvalue of the Jacobian matrix $DF_{\text{unif}}$, we make a small modification to the definition of the dominant eigenvalue of the Jacobian which will dictate the stability of the uniform state. In particular, we will consider the largest eigenvalue of $DF_{\text{unif}}$ excluding the trivial eigenvalue.

\begin{definition}[Dominant Eigenvalue of the Jacobian of a Nonlinear Random Walk]\label{def:03:01}
Consider a nonlinear random walk defined in Def.~\ref{def:02:02} with exponential bias function $f$ on an undirected, regular network. Denote the dominant eigenvalue of the Jacobian $DF_{\text{unif}}$ of the uniform state as the largest eigenvalue of $DF_{\text{unif}}$ excluding the single trivial eigenvalue $\lambda_{\text{triv}}=1$ associated with the constant eigenvector $\bm{v}_{\text{triv}}\propto\bm{1}$.
\end{definition}

We now conclude by presenting a result quantifying the dominant eigenvalue of the Jacobian $DF_{\text{unif}}$.

\begin{corollary}[Dominant Eigenvalue of the Uniform State]\label{cor:03:05}
Consider a nonlinear random walk defined in Def.~\ref{def:02:02} with exponential bias function $f$ on an undirected, regular network. The dominant eigenvalue of $DF_{\text{unif}}$is given by
\begin{align}
\lambda_{D} = \frac{\lambda}{k} - \alpha\frac{\lambda^2}{k^2} +\alpha,~~~\text{where}~~~\lambda=\text{arg\,max}_{\lambda\in\sigma'(A)}\left|\frac{\lambda}{k} - \alpha\frac{\lambda^2}{k^2} +\alpha\right|,
\end{align}
where $\sigma'(A)$ denotes the eigenvalue spectrum of the adjacency matrix $A$ excluding the eigenvalue $\lambda_{\text{triv}}=k$ that is associated with the constant vector $\bm{v}_{\text{triv}}=\bm{1}$.
\begin{proof}
A straight forward calculation shows that $A\bm{1}=k\bm{1}$, i.e., $k$ is an eigenvalue of $A$ with associated eigenvector $\bm{1}$, and this eigenvalue-eigenvector pair maps directly to the trivial eigenvalue-eigenvector pair of $DF_{\text{unif}}$. The result then follows directly from application of Theorem~\ref{thm:03:03} and Definition~\ref{def:03:01}.
\end{proof}
\end{corollary}

With these stability results in hand, we next turn to exploring the bifurcations of the uniform state and the ensuing steady-state dynamics.

\subsection{Bifurcations and Numerical Simulations}\label{sec:03:02}

\begin{figure}[t]
\centering
\epsfig{file =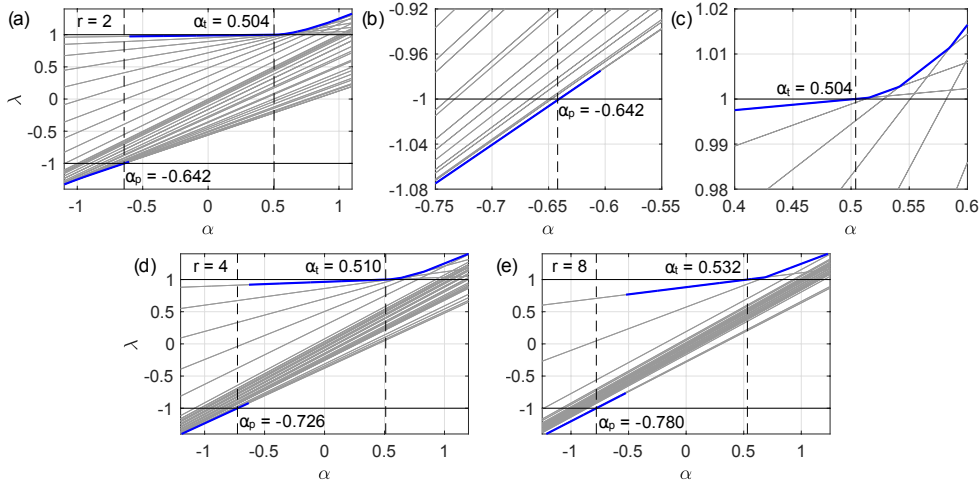, clip =,width=1.0\linewidth }
\caption{{\it Stability of the uniform state on ring networks.} Eigenvalue spectrum of the Jacobian matrix as a function of the bias parameter $\alpha$ for the uniform state for ring networks with connectivity radii (a) $r=2$, (d) $r = 4$, and (e) $r=8$. The full spectrum is plotted in as grey curves with the dominant eigenvalue highlighted in a thick, blue curve. In (b) and (c) we show a zoomed-in view near the bifurcations at negative and positive $\alpha$, respectively for the $r=2$ case.} \label{fig2}
\end{figure}

With the theoretical results from the previous subsection in hand, we now turn our eye towards exploring the bifurcations of the uniform state that occur when stability is lost. We begin by considering the family of ring networks previously described and used in Figure~\ref{fig1} and investige the spectral properties of the uniform state as the bias parameter $\alpha$ is varied. In Figure~\ref{fig2} we plot the full eigenvalue spectrum as a function of $\alpha$ of three ring networks with $N=64$ nodes with connectivity radii $r=2$, $r=4$, and $r=8$ in panels (a), (d), and (e), respectively, highlighting the dominant eigenvalue in a thick blue curve. Since the uniform state loses stability when the dominant eigenvalue exceeds one in magnitude we indicate the values $\lambda=\pm1$ with horizontal black lines and denote the critical values of the bias parameter $\alpha$ where the dominant eigenvalue crosses $1$ or $-1$ with vertical dashed lines, implying the occurrence of a bifurcation where stability of the uniform state is lost. 

First, we note that for all three cases there is a region of stability where all non-trivial eigenvalues are less than one in magnitude for sufficiently small $|\alpha|$, and that this region includes $\alpha=0$. Note that for $\alpha=0$ we recover the classical linear random walk, where for any ring network with $r\ge2$ there is a globally attracting fixed point. Second, as $\alpha$ is decreased there is bifurcation that occurs as the dominant eigenvalue crosses $-1$, indicating a period-doubling bifurcation at a critical value we denote $\alpha_p$. For the three cases of $r=2$, $4$, and $8$, this period-doubling bifurcation occurs at $\alpha_p= -0.642$, $-0.726$, and $-0.780$, respectively. In panel (b) we show a zoomed-in view of the dominant eigenvalue crossing $\lambda=-1$ at $\alpha_p$ for the $r=2$ case. Third, as $\alpha$ is increased there is another bifurcation that occurs as the dominant eigenvalue crosses $+1$, indicating a transcritical bifurcation at a critical value we denote $\alpha_t$. For the three cases of $r=2$, $4$, and $8$, this transcritical bifurcation occurs at $\alpha_t\approx 0.504$, $0.510$, and $0.532$, respectively. In panel (c) we show a zoomed-in view of the dominant eigenvalue crossing $\lambda=1$ at $\alpha_t$ for the $r=2$ case. As we'll show in Section~\ref{sec:05} this transcritical bifurcation is subcritical, giving rise to hysteresis and multistability.

\begin{figure}[t]
\centering
\epsfig{file =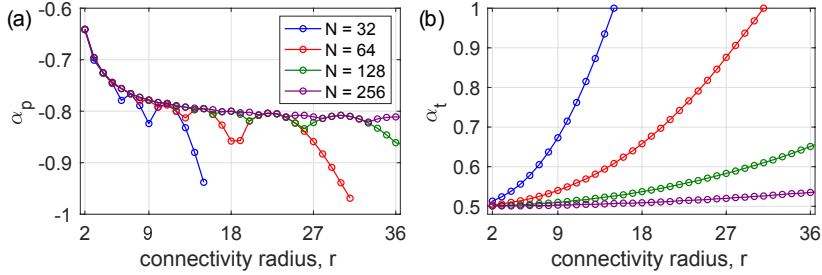, clip =,width=0.85\linewidth }
\caption{{\it Bifurcations of the uniform state on ring networks.} (a) Period-doubling and (b) transcritical bifurcation values $\alpha_p$ and $\alpha_t$ as a function of connectivity radius $r$ for ring sizes $N=32$, $64$, $128$, and $256$ (blue, red, green, and purple, respectively).} \label{fig3}
\end{figure}

Next we investigate these bifurcations in greater detail by considering variable connectivity radii and ring sizes. In Figures~\ref{fig3}(a) and (b) we plot the period-doubling and transcritical bifurcation values $\alpha_p$ and $\alpha_t$, respectively, as a function of the connectivity radius $r$ for rings of sizes $N=32$, $64$, $128$, and $256$ in blue, red, green, and purple circles, respectively. Beginning with the period-doubling bifurcations described in panel (a), we observe a complicated, non-monotonic relationship between $\alpha_p$ and $r$ for all ring sizes. (Note that for a given ring size $N$, the connectivity radius $r$ may be no larger than $N/2-1$.) In panel (b), however, we find that the relationship between the transcritical bifurcation $\alpha_p$ and $r$ is monotonic and it appears that $\alpha_t$ takes values no larger than 1.

\begin{figure}[t]
\centering
\epsfig{file =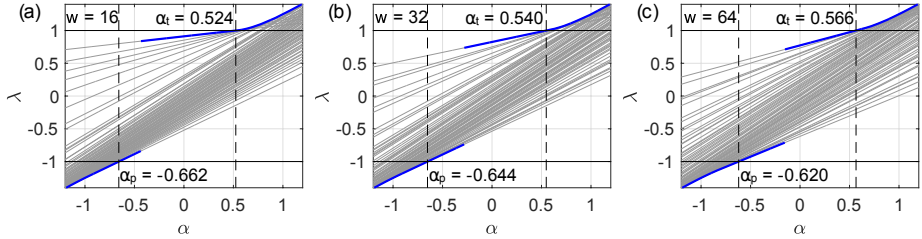, clip =,width=0.95\linewidth }
\caption{{\it Stability of the uniform state on small-world networks.} Eigenvalue spectrum of the Jacobian matrix as a function of the bias parameter $\alpha$ for the uniform state for small-world networks with connectivity radius $r=4$ and (a) $w=16$, (b) $32$, and (c) $64$ rewires. The full spectrum is plotted in as grey curves with the dominant eigenvalue highlighted in a thick, blue curve.} \label{fig4}
\end{figure}

We now move to a different family of networks that generalizes the structure of the family of ring networks considered above. In particular, we consider a variant of small-world networks~\cite{Watts1998Nature} that preserves the regularity of network structure as links are rewired. In particular, each network begins as a ring network of $N$ nodes and connectivity radius $r$. We then rewire $w$ pairs of existing links as follows. For each pair of links, $(i_1,j_1)$ and $(i_2,j_2)$, we rewire them to instead connect $(i_1,j_2)$ and $(i_2,j_1)$. Note that this particular rewiring of pairs of links maintains the regularity of the network so each node still has degree $2r$. In Figure~\ref{fig4} we plot the eigenvalue spectrum of the uniform state on small-world networks of size $N=64$ with connectivity radius $r=4$ and $w=16$, $32$, and $64$ rewires in panels (a), (b), and (c), respectively. While the overall properties of the eigenvalue spectra are similar to those of the ring networks above, we do observe a tendency for the randomness in the rewiring process to shift alter the critical values $\alpha_p$ and $\alpha_t$ denoting the period-doubling and transcritical bifurcations. In particular, for the three cases we find $\alpha_p=-0.662$, $0.644$, and $-0.620$, respectively, and $\alpha_t=0.524$, $0.5440$, and $0.566$, respectively.

\begin{figure}[t]
\centering
\epsfig{file =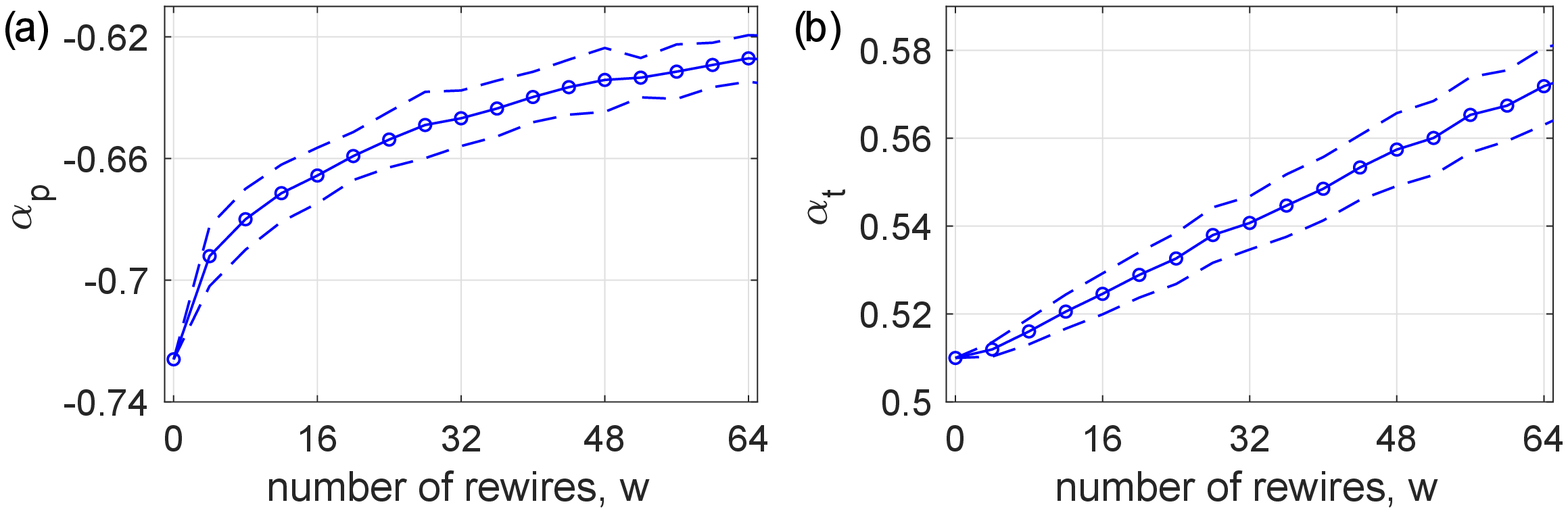, clip =,width=0.85\linewidth }
\caption{{\it Bifurcations of the uniform state on small-world networks.} (a) Period-doubling and (b) transcritical bifurcation values $\alpha_p$ and $\alpha_t$ as a function of the number $w$ of rewires on small-world networks of size $N=64$ with connectivity radius $r=4$. Each data point represents a mean over $100$ network realizations and dashed curves indicate one standard deviation.} \label{fig5}
\end{figure}

To investigate the effect of rewiring in nonlinear random walks on small world networks more closely, we plot in Figures~\ref{fig5}(a) and (b) the period-doubling and transcritical bifurcation values $\alpha_p$ and $\alpha_t$, respectively, as a function of the number $w$ of rewires in small-world networks of size $N=64$ with connectivity radius $r=4$. Each data point is a mean over $100$ network realizations and one standard deviation is indicated by dashed curves. Overall, we see that the randomness of rewirings tends to promote the period-doubling bifurcation (i.e., bring $\alpha_p$ closer to zero) but inhibit the transcritical bifurcation occurs (i.e., increase $\alpha_t$).

\section{Oscillations and Localized Patterns}\label{sec:04}

Having explored the stability properties of the uniform state in nonlinear random walks we now turn our attention to the pattern formation phenomena that occurs beyond the bifurcations where the uniform state loses stability. In particular, we will see below that the patterns that emerge can be understood and predicted using the dominant eigenvectors of the Jacobian matrix $DF$ of the uniform state, as these eigenvectors yield the modes that grow most quickly as the uniform state is perturbed.

\begin{figure}[t]
\centering
\epsfig{file =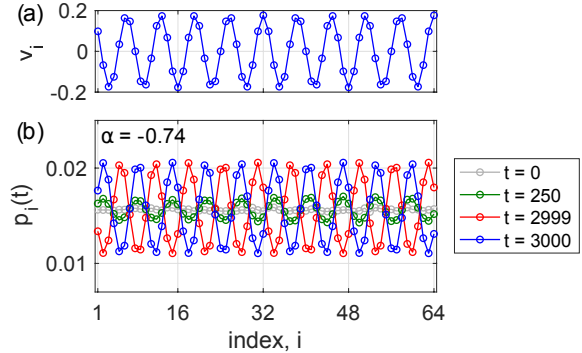, clip =,width=0.60\linewidth }
\caption{{\it Oscillating patterns on ring networks.} (a) Dominant eigenvector $\bm{v}$ of the Jacobian $DF$ of the uniform state for $\alpha = -0.74$ on a ring network of size $N=64$ with connectivity radius $r=4$. (b) Probability vector $\bm{p}(t)$ at $t=0$, $250$, $2999$, and $3000$ (grey, green, red, and blue, respectively).} \label{fig6}
\end{figure}

We begin by considering the case of negative bias and the patterns that form beyond the period-doubling bifurcation, i.e., for $\alpha<\alpha_p$. In Figure~\ref{fig1}(d) we demonstrated that for negative bias the patterns that emerge display a relatively small spatial wavelength and oscillate from one time step to another. Using a ring network of size $N=64$ and a connectivity radius of $r=4$ and setting $\alpha=-0.74$, we plot in Figure~\ref{fig6}(a) the dominant eigenvector of $DF_{\text{unif}}$ along with in panel (b) the vectors $\bm{p}(t)$ for time steps $t=0$, $250$, $2999$, and $3000$ in grey, green, red, and blue circles, respectively, indicating the initial state, a transient state, and oscillating steady-states. As the (random) perturbation evolves away from the uniform state we see that it grows nearly precisely in the shape of the dominant eigenvector and remains as such as it relaxes to the oscillating steady-state. Note that this steady-state is oscillating due to the negative dominant eigenvalue, which as discussed above, characterizes a period-doubling bifurcation. We also note that, due to the rotational symmetry of the ring structure, there is not just one dominant eigenvector, but two, corresponding to phase-shifted versions of one another, i.e., sines and cosines, with the same spatial wavelength. This is why the transient and steady-state patterns in Figure~\ref{fig6}(b) are slightly shifted from the eigenvector plotted in Figure~\ref{fig6}(a), although note that the spatial wavelength is precisely the same.

\begin{figure}[t]
\centering
\epsfig{file =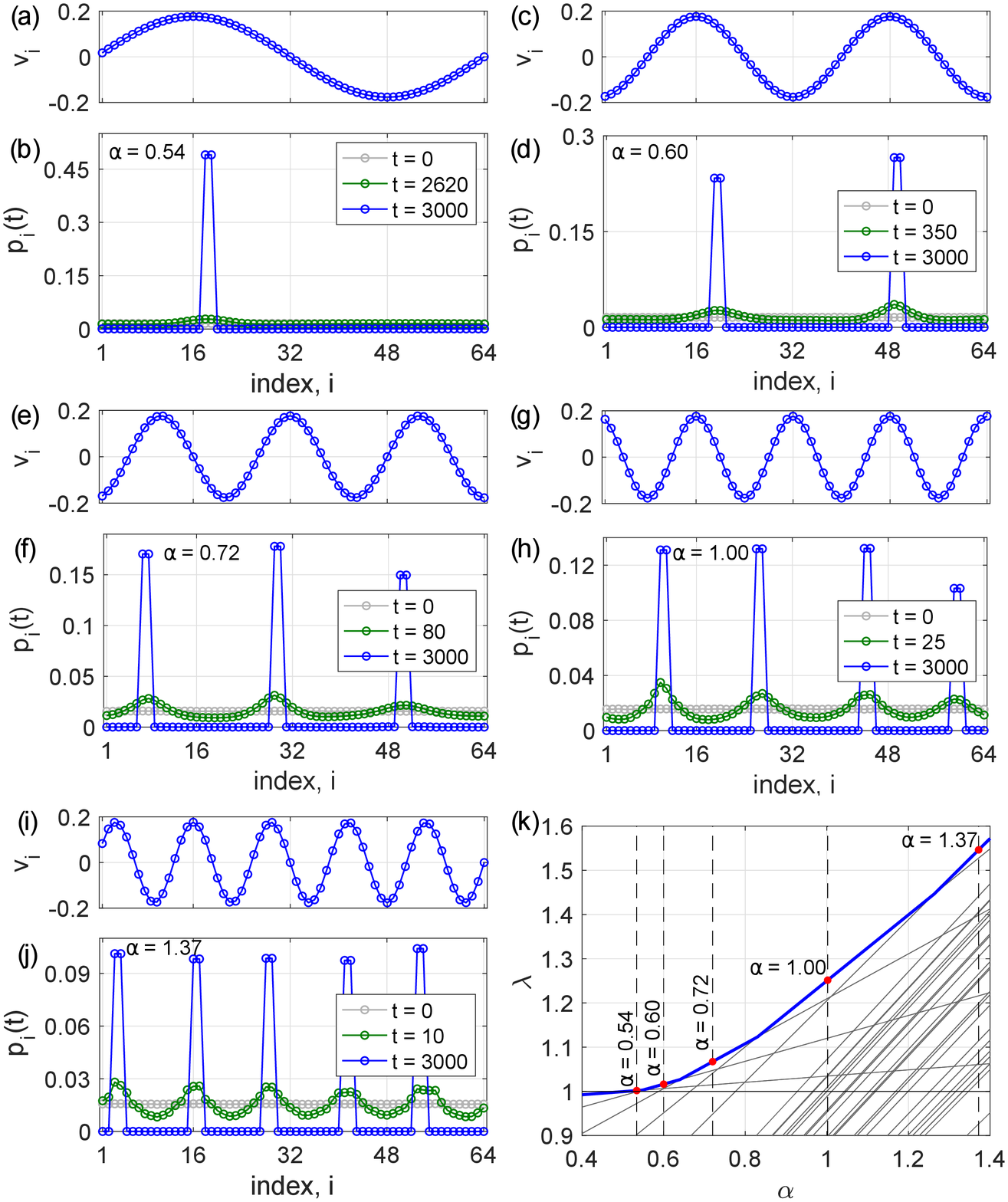, clip =,width=0.90\linewidth }
\caption{{\it Localized patterns on ring networks.} For a ring network of size $N=64$ with connectivity radius $r=4$, the dominant eigenvectors $\bm{v}$ of the Jacobian $DF$ of the uniform state and localized patterns $\bm{p}(t)$ for (a),(b) $\alpha=0.54$, (c),(d) $0.60$, (e),(f) $0.72$, (g),(h) $1.00$, and (i),(j) $1.37$. For each patter we plot the initial state, a transient state, and the steady state in grey, green, and blue circles, respectively. (k) A zoomed-in view of eigenvalue spectrum of the Jacobian as a function of $\alpha$ with the dominant eigenvalue highlighted in blue. Vertical dashed lines with red dots indicate the $\alpha$ values used in the prior panels.} \label{fig7}
\end{figure}

Next, we move to the case of positive bias, i.e., the patterns that form beyond the transcritical bifurcation characterized by $\alpha>\alpha_t$. As we demonstrated in Figure~\ref{fig1}(a)--(c), in this regime we observe a variable number of localized structures. The number of these localized structures that emerge spontaneously, it turns out, depends on the value of $\alpha$, with more localized structures emerging the further $\alpha$ is pushed beyond $\alpha_t$. Moreover, this phenomena is also dictated by the dominant eigenvectors of the Jacobian matrix of the uniform state. Using the same ring network of size $N=64$ and connectivity radius $r=4$, we plot in Figure~\ref{fig7} the dominant eigenvectors and the localized patterns that emerge from random perturbation of the uniform state for a number of $\alpha$ values: (a),(b) $\alpha=0.54$, (c),(d) $0.60$, (e),(f) $0.72$, (g),(h) $1.00$, and (i),(j) $1.37$. As noted previously, each dominant eigenvector actually has a phase-shifted counterpart (corresponding to a sine and cosine). For each pattern, we plot the initial state at $t=0$ in grey circles, a transient state at some larger time in green circles, and the steady-state reached at $t=3000$ in blue circles. First, note that for these five values of $\alpha$ we observe one, two, three, four, and five localized structures, respectively. These structures do not take the form of the dominant eigenvectors precisely, but note that for each case the number of localized structures corresponds precisely to the spatial frequency of the dominant eigenvectors. Moreover, for each case the transient state (green circles) demonstrates growth of the perturbation in a similar shape as the dominant eigenvector. Once this perturbation grows sufficiently large the nonlinearity of the positive bias localizes each peak, yielding the same number of localized structures as peaks in the dominant eigenvector. Lastly, in Figure~\ref{fig7}(k) we plot a zoomed-in view of the eigenvalue spectrum of the Jacobian $DF$ for $\alpha$ beyond $\alpha_t$, highlighting the dominant eigenvalue in blue. Here we see that the nonlinear dependence of the eigenvalues of the Jacobian matrix on the eigenvalues of the adjacency matrix result in eigenvalue crossings where the role of the dominant eigenvector switches at each crossing, thereby explaining the different dominant eigenvectors that emerge in different ranges of the bias parameter $\alpha$. In particular, each of the five values of $\alpha$ use in the previous panels are chosen from different ranges in between eigenvalue crossings. These values are indicated with vertical dashed lines and a red dot.

\begin{figure}[t]
\centering
\epsfig{file =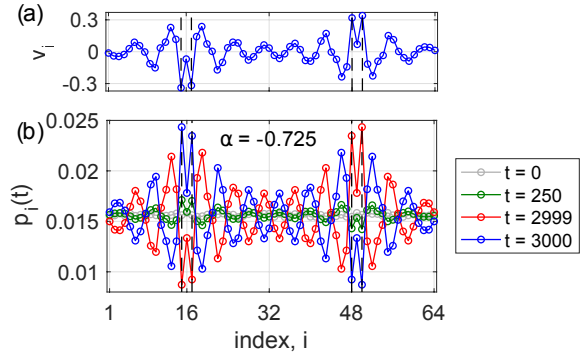, clip =,width=0.60\linewidth }
\caption{{\it Oscillating patterns on small-world networks.} (a) Dominant eigenvector $\bm{v}$ of the Jacobian $DF$ of the uniform state for $\alpha = -0.74$ on a small-world network of size $N=64$ with connectivity radius $r=4$ and a single rewire. (b) Probability vector $\bm{p}(t)$ at $t=0$, $250$, $2999$, and $3000$ (grey, green, red, and blue, respectively). Vertical dashed lines indicate nodes whose links were rewired, namely, indices $i=15$, $17$, $48$, and $50$.} \label{fig8}
\end{figure}

In addition to the patterns the form on ring networks, we also investigate the effect that rewires have on the formation of patterns using small-world networks. We consider small-world networks of size $N=64$ with connectivity radius $r=4$, and for simplicity allow a single rewiring of a pair of links. Beginning with the case of negative bias, in Figure~\ref{fig8} we plot the dominant eigenvector $\bm{v}$ and the state vector $\bm{p}(t)$ for $\alpha=-0.725$. The indices affected by the single rewiring of a pair of links are indicated using vertical dashed lines. As with the case of the ring network, the the dominant eigenvector shapes both the growth of the transient state (green circles) as well as the oscillating steady-state (red and blue circles). While this oscillating pattern also displays a relatively small spatial wavelength, the rewiring affects the amplitude of the pattern, specifically with a larger amplitude near the nodes involved in the rewiring.

\begin{figure}[t]
\centering
\epsfig{file =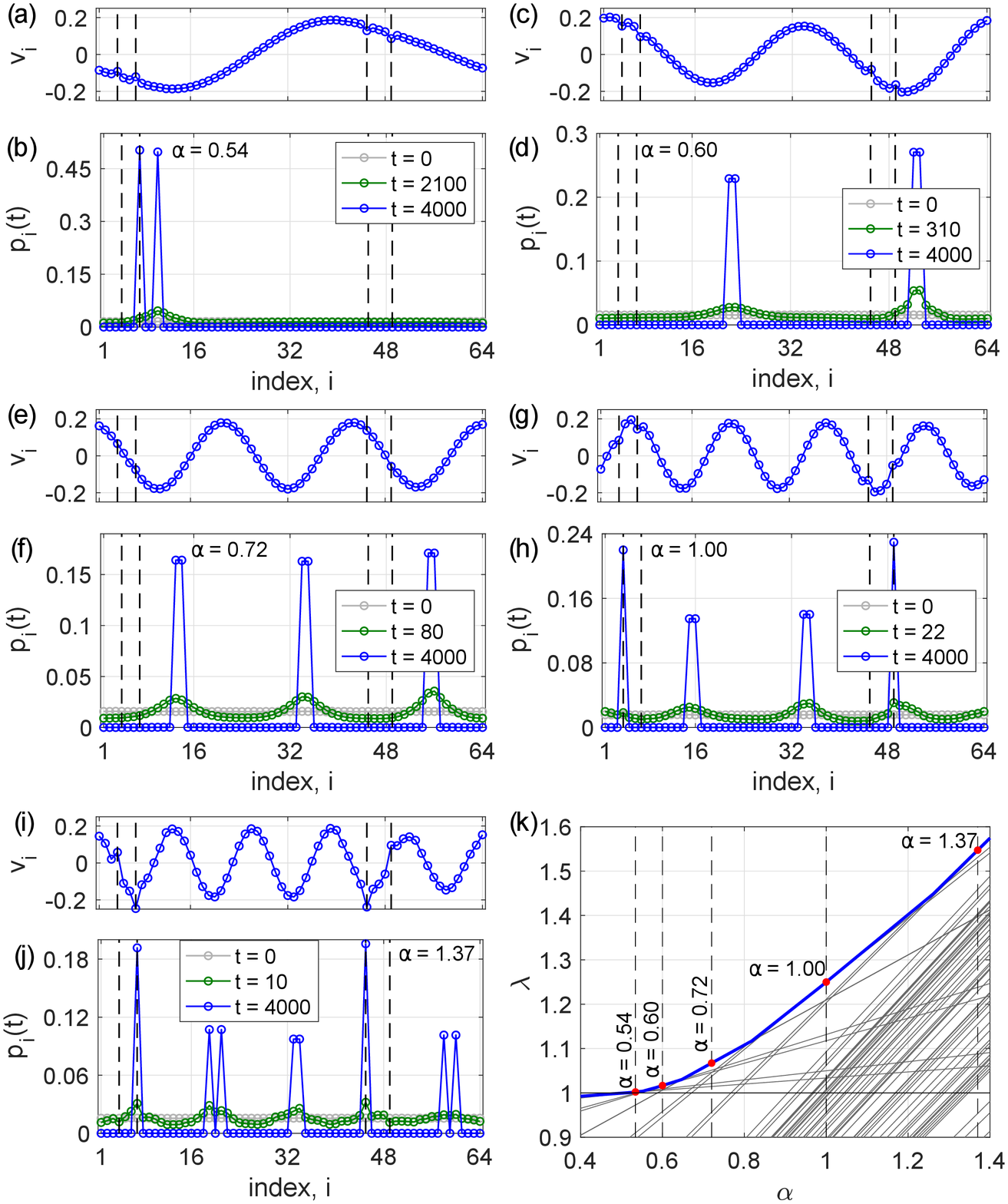, clip =,width=0.90\linewidth }
\caption{{\it Localized patterns on small-world networks.} For a small-world network of size $N=64$ with connectivity radius $r=4$ and a single rewire, the dominant eigenvectors $\bm{v}$ of the Jacobian $DF$ of the uniform state and localized patterns $\bm{p}(t)$ for (a),(b) $\alpha=0.54$, (c),(d) $0.60$, (e),(f) $0.72$, (g),(h) $1.00$, and (i),(j) $1.37$. For each patter we plot the initial state, a transient state, and the steady state in grey, green, and blue circles, respectively. Vertical dashed lines indicate the indices of nodes involved in the link rewiring. (k) A zoomed-in view of eigenvalue spectrum of the Jacobian as a function of $\alpha$ with the dominant eigenvalue highlighted in blue. Vertical dashed lines with red dots indicate the $\alpha$ values used in the prior panels.} \label{fig9}
\end{figure}

Moving to negative bias, we consider a different small-world network with the same parameter and plot in Figure~\ref{fig9} the dominant eigenvectors $\bm{v}$ and the state vectors $\bm{p}(t)$ for (a),(b) $\alpha=0.54$, (c),(d) $0.60$, (e),(f) $0.72$, (g),(h) $1.00$, and (i),(j) $1.37$. First note that, while the overall structure of the eigenvectors remains similar to those of the ring network, the rewired links cause a noticeable perturbation of the dominant eigenvectors, in particular near the nodes involved in the rewiring. The overall structure of these dominant eigenvectors being similar, the number of localized structures that emerge is still dictated by the (approximate) spatial frequency of the dominant eigenvector. The patterns the emerge, however, highlight two phenomena of note. First, as in Figures~\ref{fig9}(b) and (j), not all localized structures form at two adjacent nodes, but may occur at two nearby nodes. In general, these nodes need to be located within $r$ of one another. This may also occur in pure ring networks, but we find it occurs more often in small-world networks when the rotational symmetry of the networks is broken. Second, the precise locations of nodes involved in the rewiring do not always affect the shapes the localized patterns (in Figures~\ref{fig9}(d) and (f), for instance) but they can. In particular, in Figures~\ref{fig9}(h) and (j) we observe localized structures consisting not only of two adjacent nodes, but also even larger spikes at an isolated node. While this phenomena seems impossible at first, due to the fact that all random walkers must change locations at each time step, we note that in each case these isolated spikes occur at nodes involved in rewiring, giving them a long-range link to another node at a different party of the ring.

\section{Hysteresis and Multistability}\label{sec:05}

\begin{figure}[t]
\centering
\epsfig{file =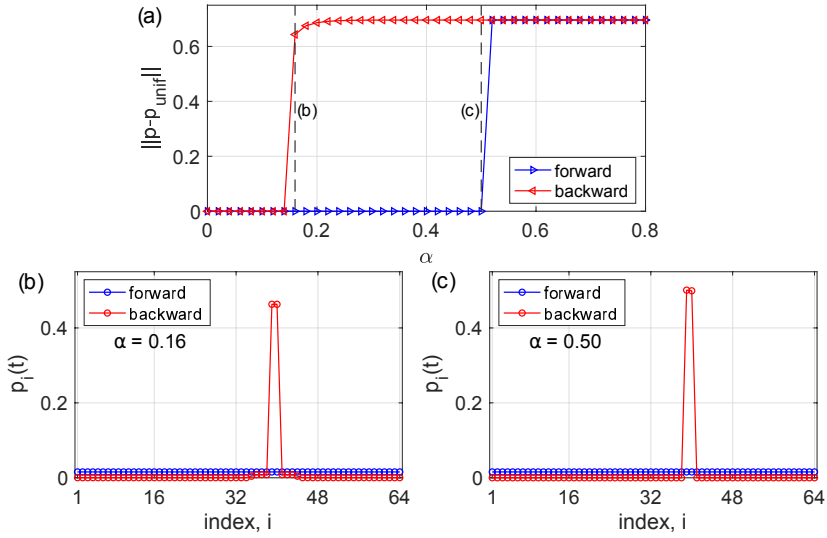, clip =,width=0.85\linewidth }
\caption{{\it Hysteresis in nonlinear random walks.} Hysteresis loop for a ring network of $N=64$ nodes with connectivity radius $r=4$ depicted using the distance from the uniform state, $\|\bm{p}-\bm{p}_{
\text{unif}}\|$, obtained by adiabatically increasing (blue arrows pointing right) then decreasing (red arrows pointing left). Example steady-state probability vectors obtained along the forward (blue) and backward (red) branches for (b) $\alpha = 0.16$ and (c) $0.50$.} \label{fig10}
\end{figure}

Before concluding, we investigate further the transcritical bifurcation characterizing the loss of stability of the uniform state for positive bias that occurs at $\alpha=\alpha_t$. Note first that after the period doubling bifurcation at $\alpha=\alpha_p$ that occurs for negative bias, the oscillating patterns that emerge take a relatively small amplitude whose shape is proportional to the dominant eigenvector, see Figures~\ref{fig6}(b) and \ref{fig8}(b). After the transcritical bifurcation, however, the localized patterns that emerge represent a larger deviation from the uniform state. This relatively large jump suggests the possibility of subcriticality in the transcritical bifurcation. We investigate this numerically using a ring network of $N=64$ nodes with connectivity radius $r=4$ by tracking the steady-state solutions that emerge as $\alpha$ is adiabatically first increased from $0$ to $0.8$, then decreased back to $0$. Specifically, beginning at $\alpha=0$ we simulate the dynamics until steady-state ($5\times10^3$ iterations), then increase $\alpha$ by a small amount, and repeat until we reach $\alpha = 0.8$. We then repeat the same process, just with decreasing $\alpha$. At each small change of $\alpha$ a small perturbation is applied to $\bm{p}(t)$. In Figure~\ref{fig10}(a) we plot the normed different of the steady-state, $\|\bm{p}-\bm{p}_{\text{unif}}\|$, as a function of $\alpha$, denoting the branches obtained by increasing and decreasing $\alpha$ with blue right-pointing arrows and red left-pointing arrows, respectively. First, along the forward branch (blue right arrows) we observe an apparently discontinuous jump from $\|\bm{p}-\bm{p}_{\text{unif}}\|=0$ to $\|\bm{p}-\bm{p}_{\text{unif}}\|=0.6960$ at approximately $\alpha=0.50$. However, along the backward branch (red left arrows) a discontinuous jump from $\|\bm{p}-\bm{p}_{\text{unif}}\|=0.6433$ to $\|\bm{p}-\bm{p}_{\text{unif}}\|=0$ occurs at the much smaller value of $\alpha=0.16$, revealing a substantial region of multistability via a hysteresis loop where both the uniform state and a localized state are both stable to perturbation. In Figures~\ref{fig10}(b) and (c) we plot the steady-states $\bm{p}(t)$ obtained at $\alpha=0.16$ and $0.50$ (towards the edges of the hysteresis loop) along the forward and backward branches (blue and red circles, respectively). These values of $\alpha$ are indicated in panel (a) with vertical dashed lines. The steady-states for these values of $\alpha$ on the forward branches are given by the uniform state and the localized states on the backwards branch are very similar, except for a small increase in the values of $\bm{p}(t)$ near the localized structure for $\alpha=0.16$. Overall, the appearance of the hysteresis loop indicates that the transcritical bifurcation at which point the uniform state loses stability is in fact subcritical.

\section{Discussion}\label{sec:06}

In this paper we have studied the pattern formation of nonlinear random walks on regular networks. Beginning with the uniform state, we have shown that this state is always a fixed point of a nonlinear random walk on a regular network topology, then presented a number of stability results for the uniform state. In particular, we link the spectral properties of the Jacobian of the uniform state to the the spectral properties of the adjacency matrix, allowing us to write the eigenvalues of the Jacobian, which dictate the stability of the uniform state, as a nonlinear function of the eigenvalues of the adjacency matrix along with the network degree and bias parameter. Exploring both ring and small-world topologies, we next identify period-doubling and transcritical bifurcations sufficiently negative and positive bias, respectively. Beyond these bifurcations we observe, for negative and positive bias, respectively, oscillating short-waverlength patterns and localized structures, respectively, whose structure can be understood using the spectral properties of the Jacobian of the uniform state, in particular the dominant eigenvector. Most notably, beyond the transcritical bifurcation for positive bias the patterns that emerge consist of a number of localized structures that depend on the bias parameter, in particular the spatial frequency of the dominant eigenvector. Lastly, we uncover a hysteresis loop for positive bias where both localized patterns and the uniform state are stable to perturbation.

In order to identify the most fundamental properties of pattern formation in nonlinear random walks, in this work we have utilized regular network topologies. However, localized and oscillating patterns are likely to emerge more generally in heterogeneous network structure, leaving their bifurcations and dynamic properties an important topic for future research. Similarly, the influence of other bias functions (here we have used a simple exponential bias function), both monotonic and non-monotonic, on pattern formation and dynamics overall in nonlinear random walks represents a largely unexplored topic for future work.


%

\bibliographystyle{siam}
\bibliography{Pattern}

%
\end{document}